\documentclass[article]{elsarticle}

\usepackage{lineno,hyperref}
\usepackage[utf8]{inputenc}
\usepackage{xcolor}
\usepackage[colorinlistoftodos]{todonotes}
\usepackage[normalem]{ulem}
\usepackage{amssymb,amsmath}

\modulolinenumbers[5]

\journal{Journal of \LaTeX\ Templates}

\begin{document}

\begin{frontmatter}

\title{Empirical determination of the optimum attack for fragmentation of modular networks}
%\title{Elsevier \LaTeX\ template\tnoteref{mytitlenote}}



\author[mymainaddress]{Carolina de Abreu}
\author[mysecondaryaddress]{Bruno Requião da Cunha}
\author[mymainaddress]{Sebastián Gonçalves\corref{mycorrespondingauthor}}
\cortext[mycorrespondingauthor]{Corresponding author}
\ead{sgonc@if.ufrgs.br}

%\author[mysecondaryaddress]{Sebastián Gonçalves\corref{mycorrespondingauthor}}
%\ead{sgonc@if.ufrgs.br}

\address[mymainaddress]{Instituto de Física, Universidade Federal do
  Rio Grande do Sul, 91501-970 Porto Alegre, Rio Grande do Sul,
  Brazil} \address[mysecondaryaddress]{Superintendência da
  Polícia Federal no Rio Grande do Sul, 90160-093, Porto Alegre, Rio
  Grande do Sul, Brazil}

\begin{abstract}
All possible removals of $n=5$ nodes from networks of size $N=100$ are
performed in order to find the optimal set of nodes which fragments
the original network into the smallest largest connected
component. The resulting attacks are ordered according to the size of
the largest connected component and compared with the state of the art
methods of network attacks.  We chose attacks of size $5$ on relative
small networks of size $100$ because the number of all possible
attacks, ${100}\choose{5}$ $\approx 10^8$, is at the verge of the
possible to compute with the available standard computers.  Besides,
we applied the procedure in a series of networks with controlled and
varied modularity, comparing the resulting statistics with the effect
of removing the same amount of vertices, according to the known most
efficient disruption strategies, {\em i.e.}, High Betweenness Adaptive
attack (HBA), Collective Index attack (CI), and Modular Based Attack
(MBA).  Results show that modularity has an inverse relation with
robustness, with $Q_c \approx 0.7$ being the critical value. For
modularities lower than $Q_c$, all heuristic method gives mostly the
same results than with random attacks, while for bigger $Q$, networks
are less robust and highly vulnerable to malicious attacks.

%High Betweenness Adaptive (HBA),
%Collective Index (CI), and Module-Based (MBA) prescriptions. 
\end{abstract}

\begin{keyword}
modular networks\sep network robustness \sep attacks
\MSC[2010] 05C85\sep 68R10\sep 90C35
\end{keyword}
\end{frontmatter}

%\linenumbers

\section{Introduction}
%literature review
Robustness of complex networks is a very important issue in a wide
range of distinct applied disciplines such as physics, biology,
engineering, sociology and criminology or police science
\cite{jingfang, muro, BELLINGERI2018316, Hu201710547, daCunha2018}--
just to name a few.  Network theory provides the natural framework for
the systematic study of the robustness/weakness of interconnected
systems; in particular, to help identify the key topological factors
which play a crucial role in the cohesion of a networked system.  For
instance, scale-free graphs are fragile to the targeted removal of its
highest degree (hubs) nodes, while random graphs loose its cohesion
after the random deletion of a critical fraction of vertices. However,
the problem of finding the minimal set of nodes that if removed would
break the network into non-extensive components is
NP-hard~\cite{Morone:2015aa} and has no general analytic solution.
Indeed, it is a combinatorial complexity problem, for if we want to
know which set of $n$ nodes produces the maximum damage in a network
of size $N$, all possibilities must be tried which is a total of
${N}\choose{n}$ sets.  Even for relatively small networks, the problem
could be computationally intractable depending on the number of nodes
that we want to remove. For example, checking the effect of removing
$10$ nodes from a small network of size $50$ gives $\approx 10^{10}$
different sets to verify which one is the optimum.  Therefore, many
suboptimal methods have been proposed in
%to get as close as possible to that optimal set of nodes
the last few years~\cite{lozano2017optimizing, tian2017articulation, ren2018generalized,
kitsak2010identification, geisberger2008better},
they are based on different heuristic rankings and are
computationally efficient.  Morone and Makse, for example,
have shown that the so called ``weak nodes'' identified by the
Collective Influence algorithm (CI), \emph{i.e.} the ones with
low-degree surrounded by hierarchical coronas of hubs, are powerful
influencers~\cite{Morone:2016aa}. However, many networks tend to group
into modular structures ---clusters densely connected internally but
weakly linked among them---, in a way that affects the network
robustness.  Such was shown recently for modular
networks~\cite{Cunha:2015aa}, pointing out that bridges connecting
communities are more prominent to network cohesion than hubs or Morone
and Makse's ``weak'' nodes.  Moreover, the assumption of local
tree-like structure, central in Morone and Makse algorithm, does not
apply for modular networks. In this sense, Kobayashi and
Masuda~\cite{Kobayashi:2016aa} combined the Morone-Makse algorithm and
coarse graining to propose a method of fragmenting networks by
targeting collective influencers at mesoscopic level, proving to
outperform the original algorithm for modular networks at a reasonable
computational cost.

Modular networks are quite different from pure random or small-world
networks for example.  Indeed, the modularity feature of a given
network signals the presence of other than pure random connection
among nodes; something inherent of social networks for instance, where
individual tend to form communities. Besides, modularity is known to 
play a key role in a wide variety of phenomena, from crime~\cite{daCunha2018} 
to brain networks~\cite{brain} for example.

Regarding robustness, interdependent modular networks are known to be
much more vulnerable than typical graphs, and localized disruptions
can give rise to powerful avalanches in power grids, biological, and
financial networks~\cite{Shekhtman:2016aa, shekhtman2018spreading}.
Phase transitions in modular networks from a connected to a dismantled
phase are more abrupt than in typical graphs.  Moreover, Shekhtman
\emph{et al.} developed a theoretical framework for network disruption
of modular networks showing that such systems might undergo either a
double first-order phase transition (inter and intra-module
fragmentations when there are fewer modules densely intra-connected)
or a single one (when there are many modules
interconnected)~\cite{Shekhtman:2015aa}.

%Faqeeh \emph{et al} have shown that several coexisting percolating clusters can emerge in networks with finite and sufficiently small number of interlinks between modules and that modular message passing has an important role in quantifying epidemic outbreaks in SIR models of disease spread.
%Emergence of coexisting percolating clusters in networks
%Ali Faqeeh,1 Sergey Melnik,1 Pol Colomer-de-Simo ́n,2 and James P. Gleeson

Noteworthy, Module-Based (MBA) and High
Betweenness Adaptive (HBA) attacks are known to atomize modular
networks with the least number of removals, with MBA being
computationally much more feasible~\cite{Perfo}.  However, it is not clear yet how close are these methods to the optimal fragmentation point, which is the largest connected component of minimum size that results after $n$ nodes are removed.
%~\footnote{By optimal fragmentation point of a network we refer to the largest connected component of minimum size that results after $n$ nodes are removed.} 
Indeed, it is not known from what value of modularity the modular property stands out, for example. 
This is precisely what we address in the present contribution. In order
to do that, we perform a brute-force computational analyses, generating all ${N}\choose{n}$
possible lists of removal, and comparing the resulting statistics with
HBA, CI, and MBA methods of attack. 

In the next section, we present the method, including the description of network benchmark, the brute force, and the heuristic method of attacks. Then the results are presented with the statistic of brute force attacks compared with the heuristic procedures. The final section is devoted to conclusions.

%\paragraph{Functionality} The Elsevier article class is based on the standard article class and supports almost all of the functionality of that class. In addition, it features commands and options to format the
%\begin{itemize}
%\item document style
%\item baselineskip
%\item front matter
%\item keywords and MSC codes
%\item theorems, definitions and proofs
%\item lables of enumerations
%\item citation style and labeling.
%\end{itemize}

\section{Method}

\subsection{Networks}
In order to generate graphs exhibiting community structure and heterogeneous node degree distribution, we use the LFR Benchmark, which is an artificial graphs benchmark developed with the aim of testing community detection algorithms \cite{Lancichinetti:2008aa}.
In the LFR benchmark, the modularity of a network is controlled by the mixing parameter $\mu$ which represents the fraction of a node's links with nodes outside the community to which such node belongs.

Considering a complete partition that organizes a given network into several communities, a general expression for calculating the partition's modularity of that network is given by~\cite{Newman:2004aa}  
\begin{equation}
Q = \frac{1}{2m} \sum_{i,j} \left[ A_{ij} - \frac{k_i k_j}{2m} \right] \delta(c_i,c_j),
\end{equation}
where $A_{ij}$ is the adjacency matrix ($A_{ij}$=1 if nodes $i$ and $j$ are connected and $0$ otherwise), $k_{i}$ is the vertex degree of node $i$, $c_{i}$ is the community to which node $i$ belongs to, $m$ is the total number of edges, and $\delta(c_i,c_j)=1$ if nodes $i$ and $j$ belong to the same community and $0$ otherwise. 
The modularity $Q$ is then a scalar value between $-1$ and $1$, which points if the local link density of the subgraphs defined by this partition differs from the expected density in a randomly wired network.
Negative values of modularity mean the nodes in the subgraphs do not form communities.
If $Q$ is zero then the connectivity between the nodes inside the subgraph are random, fully explained by the degree distribution. 
When $Q>0$ the partition is said to be suboptimal up to $Q\sim 0.5$, a value which might be due to statistical fluctuations of random networks, but with no ``true" communities separation. Thence, we are interested in networks with $Q \gtrapprox 0.5$, which show optimal partition~\cite{netscibook}.

Therefore we generate 157 networks with modularity varying from $0.567$ to $0.853$, tunable by the mixing parameter $\mu$ that lies inside the interval $[0.1,0.6]$. The generated networks also show a total number of nodes $N = 100$, average degree $\langle k \rangle = 3$, and maximum degree $k_{max}= 6$. 

\subsection{Brute Force Attacks}
{\em Brute force} attacks stands for the evaluation of all possible attacks of size $n$ on a network of size $N$, {\em i.e} counting the remaining size of the biggest connected component after deleting $n$ nodes. 
As there are ${{N}\choose{n}}$ possible choices for the $n$ nodes in a network of size $N$, this is a hard problem, generally unfeasible except for limited attacks on relatively small networks.
In this contribution we chose networks of size $N=100$ over which we perform attacks removing $n=5$
%~\footnote{Even when $n=5$ could seem a very discrete attack, it represents 5\% of the nodes of a network of size $N=100$, which is a sensible attack.} 
nodes, as a sensible compromise between interesting and/or representative sizes (resembling or approaching real interesting system) and computational cost. Even when $n=5$ could seem a very discrete attack, it represents 5$\%$ of the nodes of a network of size $N=100$, which is a sensible attack.
Therefore, for each network generated, we perform an attack by removing 5 out of the 100 nodes, 
giving ${{100}\choose{5}} \approx 10^8$ of total computed attacks for each network 
tested~\footnote{The CPU time to compute the $\approx 10^8$ different attacks necessary to make 
the histogram of Figure~\ref{fig:bruteforce} is $\approx 45'$ in an Intel(R) Core(TM) i7 \@ 3.60GHz per CPU. 
For a network of size $N=200$, that time would increase to roughly one day.}. 
Since our interest is in the damage caused, for each one of those attacks, we record the size of the graph's remaining giant component, $GC$. 
Notice that after the removal of $n$ nodes from a network of size $N$, we have always $1 \leq GC \leq N-n$.

\subsection{Heuristic Methods}
In what follows, we compare the results of the empirical brute force attacks with those obtained through the application of well established heuristic methods of network fragmentation. Three methods were used for that purpose which are described bellow:

\noindent {\bf High Betweenness Adaptive Attack (HBA)}: it classifies a graph's nodes according to the betweenness centrality measure in descending order.
Betweenness centrality measure for a node $v$ can be defined as: 
\begin{equation}
g(v) = \sum_{s\neq v \neq t} \frac{\sigma_{st}(v)}{\sigma_{st}}, 
\end{equation}
where $\sigma_{st}$ is the number of shortest paths between nodes $s$ and $t$, and $\sigma_{st}(v)$ is the number of those paths that pass through node $v$. 

\noindent {\bf Module-Based Attack (MBA):}
%This method, introduced by some of us~\cite{Cunha:2015aa} some years ago, starts by extracting the modular structure of the network using a fast community detection algorithm like Louvain method for example~\cite{Blondel:2008aa}. 
%Then the nodes connecting communities are ordered by decreasing betweenness and those are the ones to be removed in such order~\footnote{In doing that, only nodes that remain in the biggest connected component and stays connected with other communities are removed, otherwise are skipped.}
This method starts by identifying the modular structure of the network through a community detection algorithm, such as the Louvain method \cite{Blondel:2008aa}. The nodes that connect communities are ordered by their betweenness centrality, so that the ones with highest betweenness are removed \cite{Cunha:2015aa}.

\noindent {\bf Collective Influence (CI):}
Removes nodes according to their Collective Influence score~\cite{Morone:2016aa}. For a node $i$, Collective Influence can be defined as:
\begin{equation}
CI_\ell(i) = (k_i - 1) \sum_{j \in \partial B(i,\ell) } (k_j - 1),
\end{equation}
where $k_j$ is the degree of the nodes in the edge of a ball of radius around node $i$, and $k_i$ is the degree of node $i$. Like HBA, it is an adaptive method.

\section{Results}
\begin{figure}[ht]
    \centering
    \includegraphics[width = 0.49\textwidth]{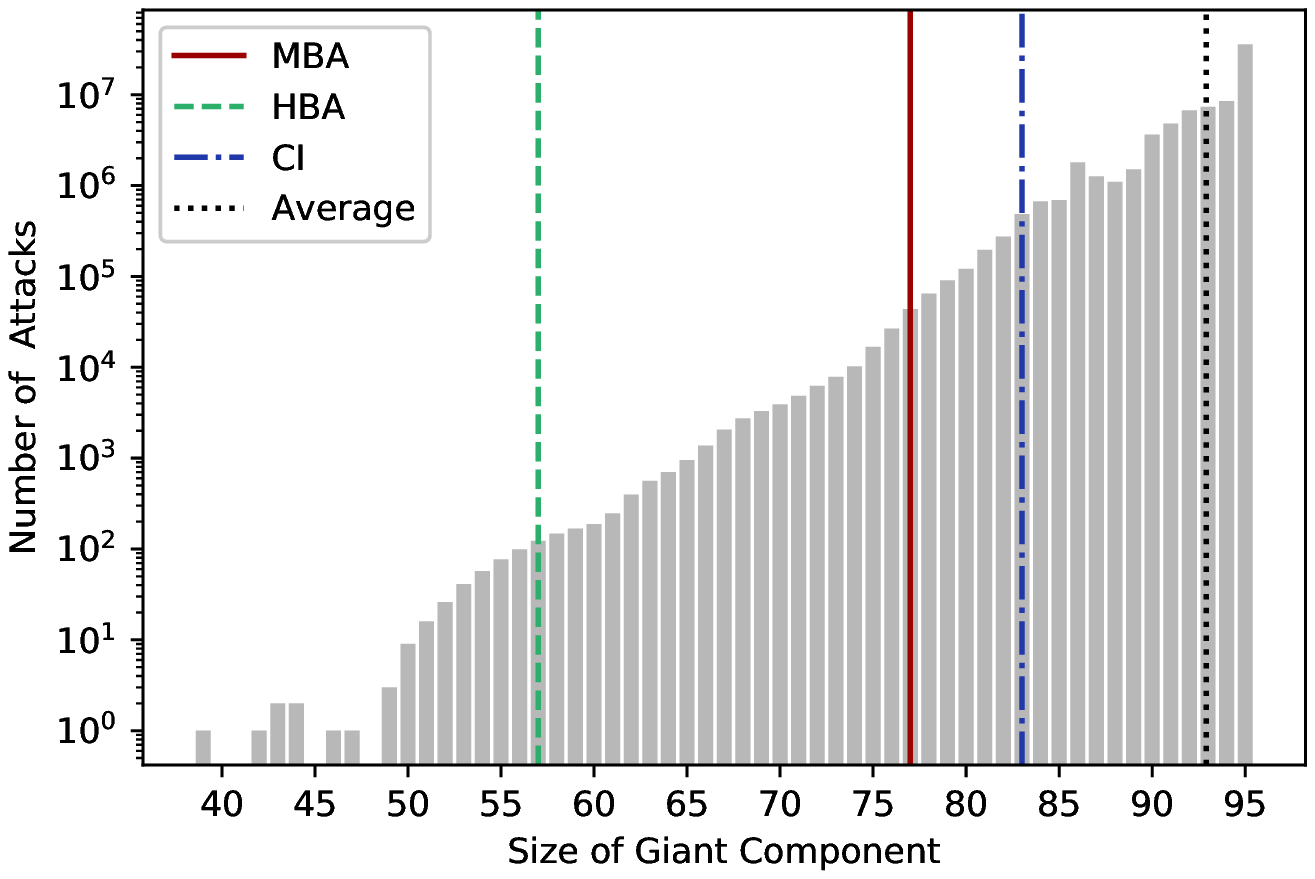}
    \includegraphics[width=0.49\textwidth]{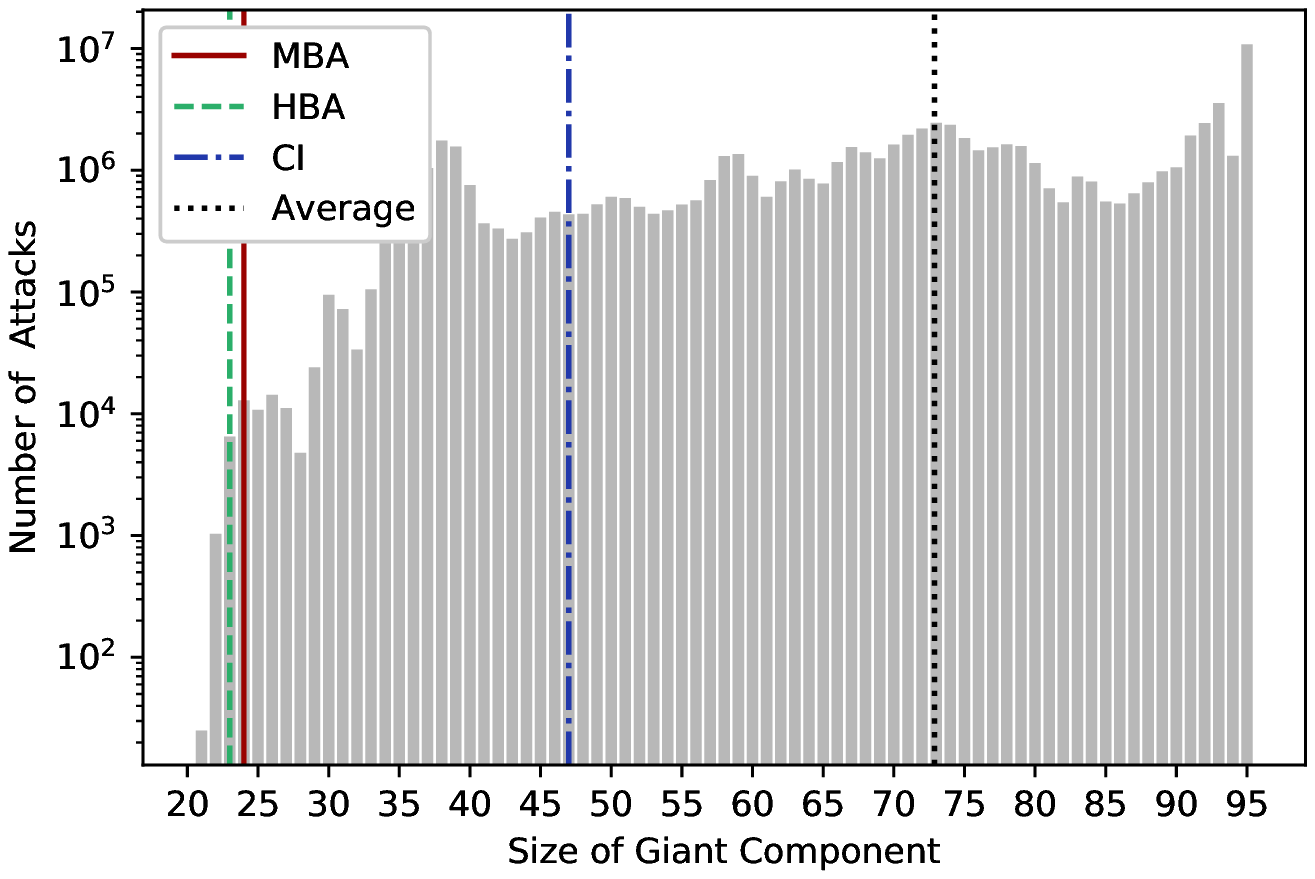}
    \caption{Histograms of all ${{100}\choose{5}}$ possible attacks of $n=5$ nodes on two networks of size $N=100$, but different modularity, $Q=0.759$ and $Q=0.808$ as indicated in the figure. The horizontal axis represents the size of the remaining giant component ($GC$) after deleting $5$ nodes and the vertical axis is the number of attacks producing the same $GC$.
    The colored vertical lines represent the value of $GC$ after the attacks according to the heuristic procedures MBA (solid red), HBA (dashed green), and CI (dot dashed blue). The black dotted line represents the expected value of $GC$.
    The optimal fragmentation points are $GC=39$ for the $Q=0.759$ network and $GC=21$ for the $Q=0.808$ network.}.
    \label{fig:bruteforce}
\end{figure}

\begin{figure}
\includegraphics[width=\textwidth]{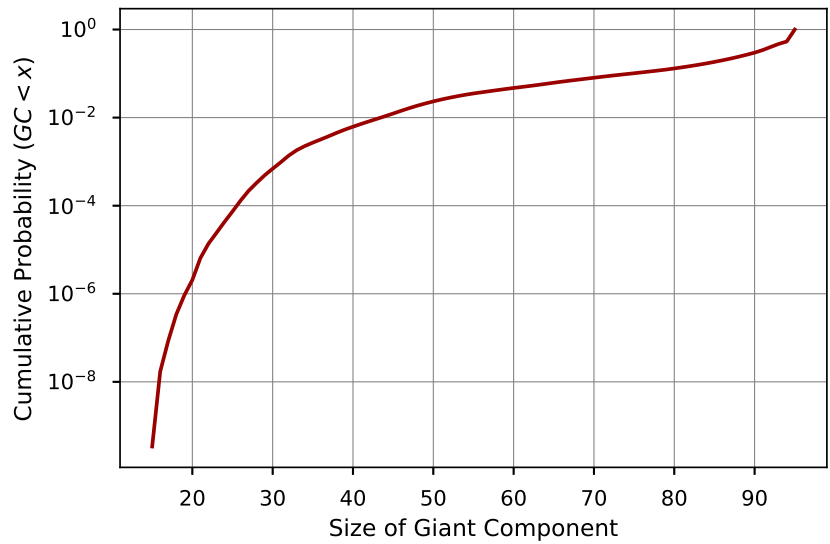}
\label{percentage}
\caption{Cumulative distribution of probabilities of $GC$ resulting after a random attack of size $n=5$ in a network of size $N=100$. The distribution is the result of all attacks over all $157$ realizations. It can be seen that while 1\% of the attacks results in a network reductions to more or less 50\% of the original size, the probability of a random attack reducing the network to a component of 20\% or less of its original size is $\approx$ 1 in a million.}
\end{figure}

The main result we present is the damage inflicted by each attack on the original graph. 
We measure the damage in terms of the size of the remaining largest connected component ($GC$).
Figure~\ref{fig:bruteforce} shows the result of the brute-force attack of size $n=5$ to two examples of networks of size $N=100$, but different modularity. Each bar represents the number (height) of attacks that reduces the network to the corresponding size (position) of the giant component.
Clearly, by performing an heuristic attack it will give much better results, in terms of damage, 
that a random selection. As removing 5\% of the nodes, in 99\% of the cases, the remaining largest 
connected component is bigger than 80\% of the original one (Figure~\ref{percentage} 
shows the probability of obtaining a certain damage in a random attack).

\begin{figure}[h!]
    \centering
    \includegraphics[width = \textwidth]{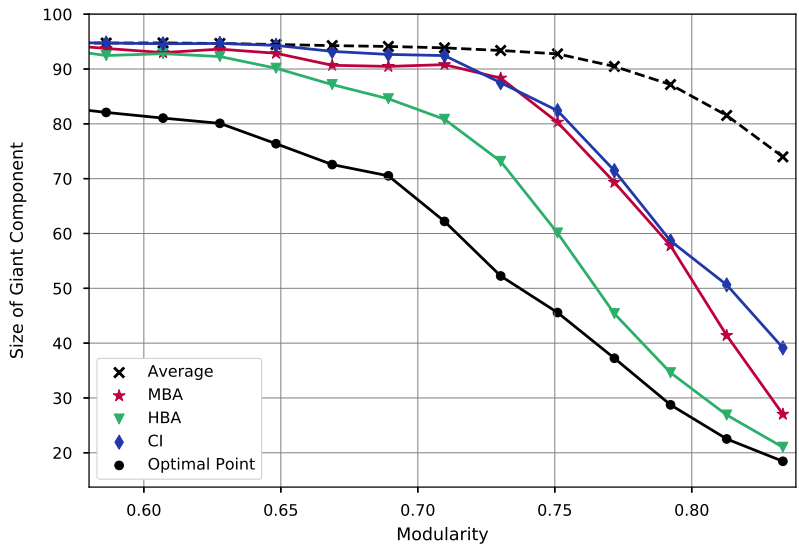}
    \caption{$GC$ after removing $5$ nodes according to the three heuristic method analyzed (MBA, HBA, and CI) vs modularity $Q$ of the networks. Each points represents the $GC$ averaged over several realization of networks for a fixed $Q$. Lower curve is the optimal attack obtained after checking all possible ones. Upper curve is the average $GC$ over all possible attacks.}
    \label{fig:gc}
\end{figure}

All the quantities presented in this contribution are averages performed over ranges of modularity in intervals of size $\Delta Q = 0.02$. Figure~\ref{fig:gc} exhibits all the results of the brute force attacks, along with the heuristic
methods results as well, as a function of the modularity. That figure condenses a good part of the results of the present contribution; it shows the remaining largest component sizes while varying the modularity, for different types of attacks. The solid black and full circles curve represents the best possible attack to a network, \emph{i.e.} the one that reduces the giant component to its lowest value. We also show the average values for a random attack which is obtained by computing the average of all possible attacks (cross and dashed black curve) of size $n=5$.
The three middle curves display the values for targeted attacks: MBA (star and red line), HBA (triangle and green line), and CI (diamond and blue line).

The figure shows that the difference between the expected value of random attacks and the optimal dismantling point increases with  modularity, setting the upper and lower limits respectively for any heuristic attack. One clear conclusion is that randomly removing a set of nodes becomes less efficient as topological communities become more prominent.
Moreover, as the modularity increases the remaining largest connected component becomes smaller and the network itself becomes less robust to any kind of attack, even a random one.

\begin{figure}
    \centering
    \includegraphics[width=\textwidth]{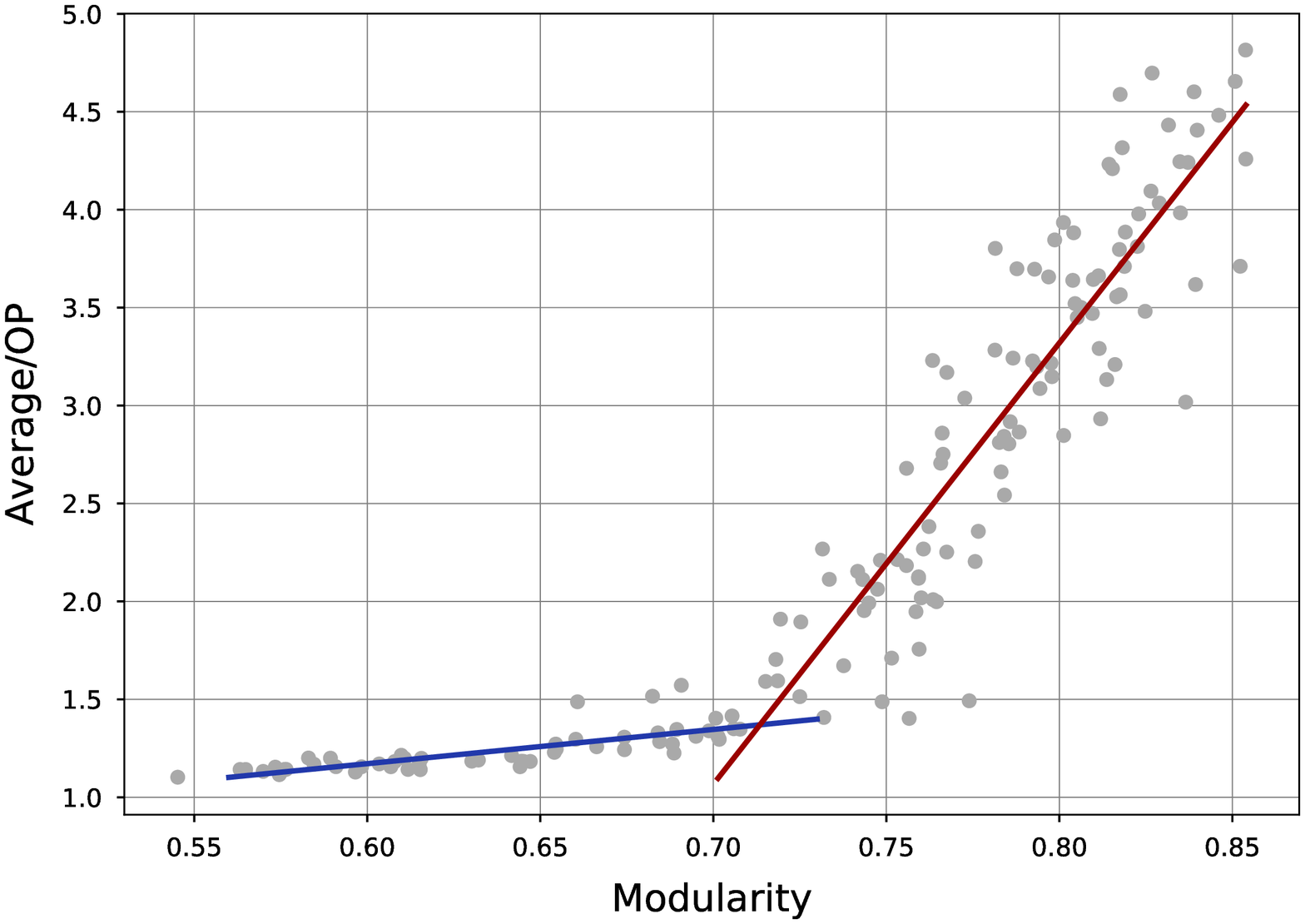}
    \caption{Ratio between the average of attacks and the optimal one as a function of modularity. Each point represents the ratio for a network realization at each value of modularity.
The blue and red lines are linear regressions of the data below  $Q=0.73$ and above $Q=0.7$ respectively, resulting in an intersection at $Q_c \approx 0.71$. Blue line: $r^2 = 0.62$, standard error $=0.19$. Red line: $r^2 = 0.83$, standard deviation $=1.0$.}
    \label{fig:linregress}
\end{figure}

In Figure~\ref{fig:linregress} we show the ratio between the average of attacks and the optimal value of the 
remaining giant component as a function of modularity. Both values, expected and optimal, are calculated 
for each network by means of the brute-force attack. Clearly there is a high correlation between these 
two quantities, as well as a rapid increase in the area of medium ($Q \approx 0.7$) to high modularity 
($Q \gtrapprox 0.7$).

\begin{figure}
    \centering
    \includegraphics[width = \textwidth]{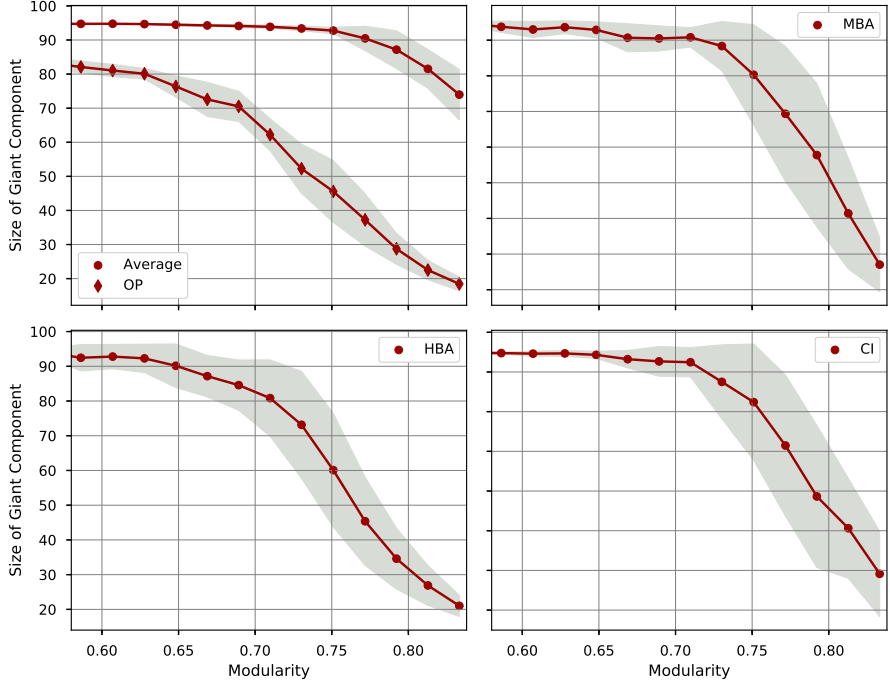}
    \caption{GC after removing 5 nodes for the three heuristic method analyzed (MBA, HBA, and CI) vs modularity Q of the networks. Each point represents the GCS averaged over bins of size $\Delta Q = 0.02$, and after applying a moving average with a window of two points. Shadow region around the curve represents the standard deviation of the result over the different realizations for each modularity. The Average curve is the average over all realization and all possible attacks for each one of them. Shadow region represents the standard deviation.}
    \label{fig:subplots}
\end{figure}

In Figure~\ref{fig:subplots} we can notice that the slope of the curves of the heuristic methods, as well the expected values curve, decrease as the modularity increases, which is expected since the number of nodes that can dismantle the network easily also increases along with the modularity. This result confirms the theoretical observation of Shekhtman \emph{et al.} that as the modularity increases networks tend to be more fragile, even facing a single fragmentation phase transition when facing failures~\cite{Shekhtman:2015aa}.

\begin{figure}[h!]
    \centering
    \includegraphics[width = 0.32\textwidth]{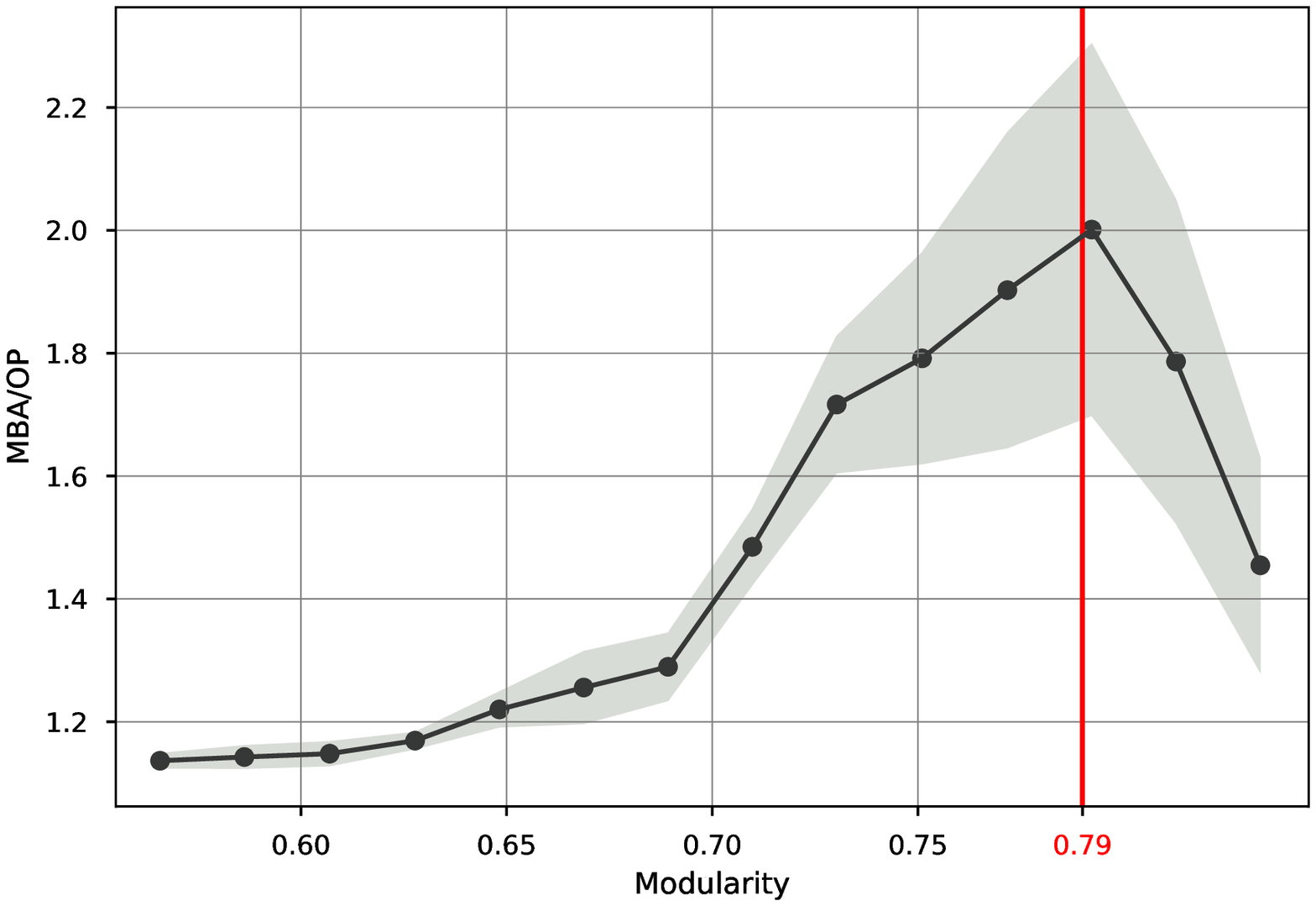}
    \includegraphics[width = 0.32\textwidth]{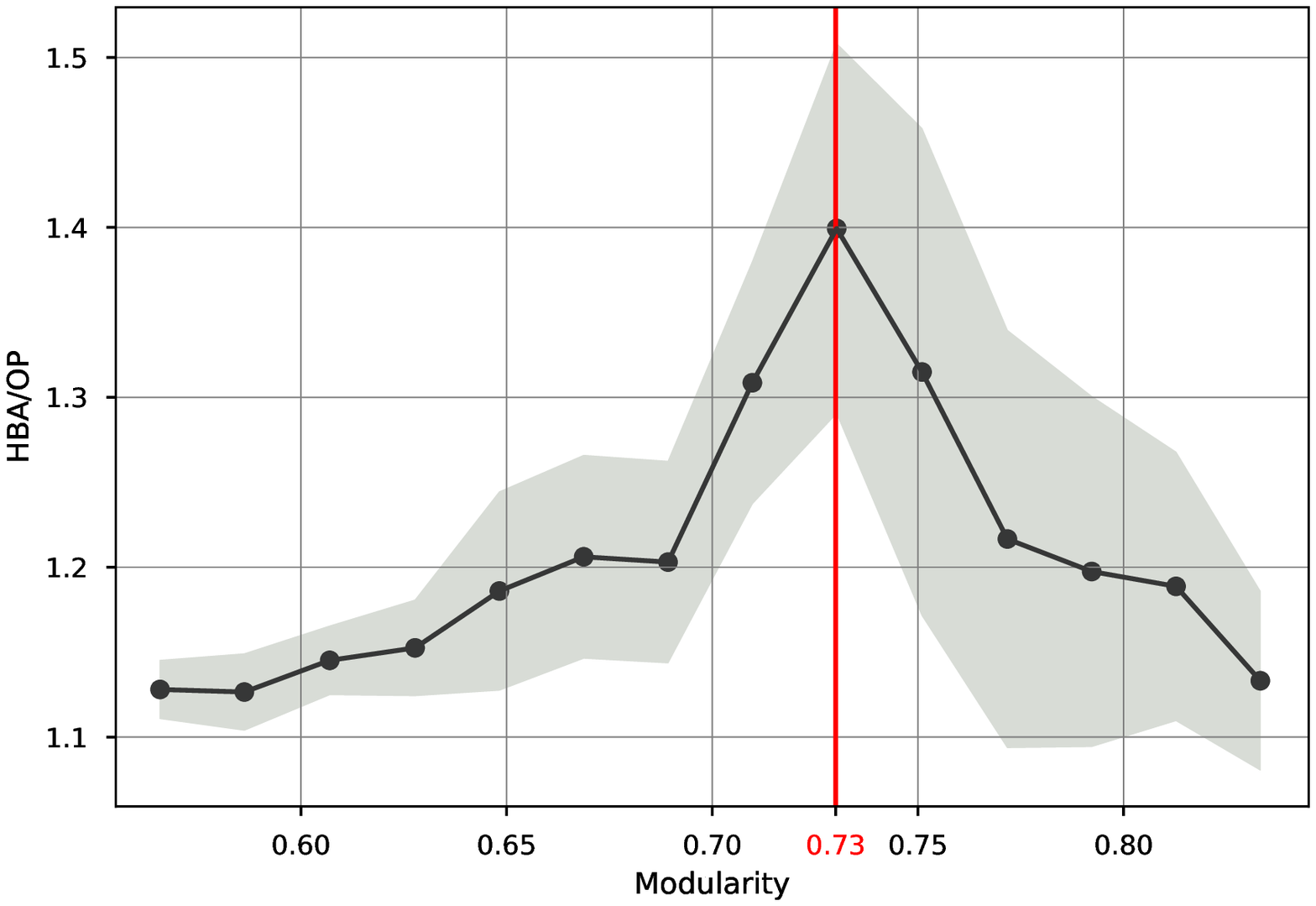}
    \includegraphics[width = 0.32\textwidth]{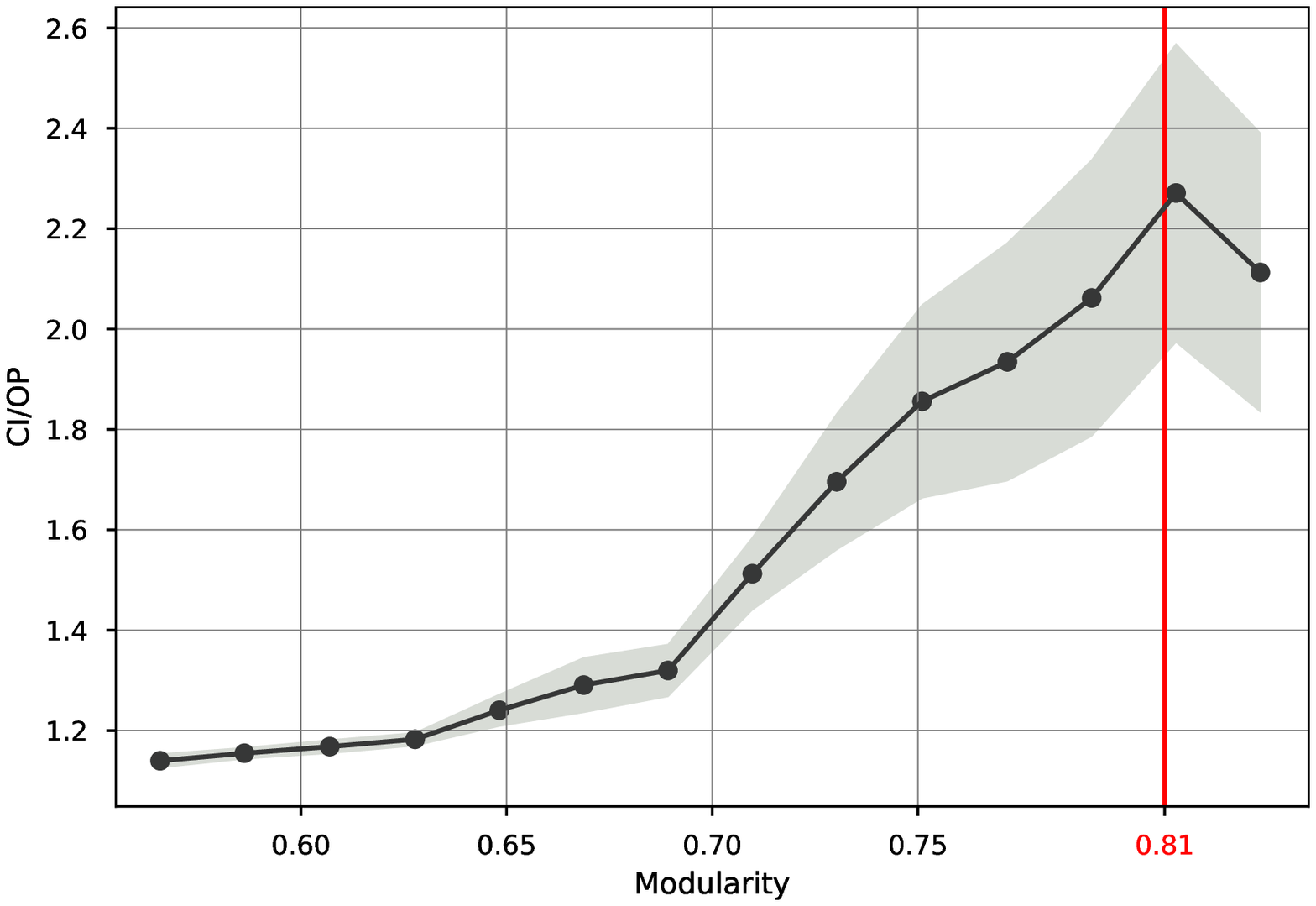}
    \caption{$GC$ ratio of the three heuristic methods relative to the optimum point as a function of modularity.
  Each point represents a different network realization for a given modularity according to the LFR 
benchmark. Squares represent the average and the line is an interpolation. Grey region represent the 
standard deviation at each modularity value. Left: MBA, Center: HBA, and Right: CI.
For each method, the red vertical line shows the local maximum, since where each heuristic strategy begins to 
re-approach the optimum fragmentation $Q^{CI}_c=0.81$, $Q^{MBA}_c=0.78$ and $Q^{HBA}_c=0.73$.}
    \label{fig:bforce-comp}
\end{figure}

% \begin{figure}
%   \centering
%   \includegraphics[width = \textwidth]{imagens/MBA-OP.png}
%   \caption{$GC$ ratio of the MBA method relative to the optimum point as a function of modularity.
%   Each point represents a different network realization fro given modularity according to the LFA benchmark. Squares represent the average and the line is an interpolation. Grey region represent the standard deviation at each modularity value. \T{Temos que padronizar. Ou se usa a razão Heur/Op ou se usa a diferença Heur-Op. Nessa figura tá a diferença, nas outras tá a razão. Eu acho melhor usar só a razão. {\R Concordo, alias algo esta errado nessa figura: $GC(MBA) - GC(OP) > 60$, para $Q=0.8$?? Não é possível! --Sebastian}}\textcolor{blue}{Essa figura tem que apagar. A única que tem que entrar é a 5 que mostra a razão entre o GC(heur)/GC(opt). E nesses casos estimar pela derivada nula dos dados.}}
%   \label{fig:MBA}
% \end{figure}

The heuristic approach to network fragmentation shows two distinct regimes. For modularities lower than a critical value $q_{c}$ (which we determine below), the heuristic attacks curves distance themselves from the optimal point. After the inflection point the difference between the size of the giant component in the optimal strategy and in the heuristic attacks starts to diminish until the MBA and HBA attacks get very close to the optimum dismantling point at $0.86$. However, for high values of modularity the difference between optimal, HBA, and MBA results is such that time complexity should be taken into account when searching for the optimal dismantling set for larger networks.  

To measure the critical value $Q_{c}$ at which the attack curves start to get closer to the optimal dismantling
set we plot the ratio of the remaining largest connected component after each attack and the optimum point 
as depicted in Figure~\ref{fig:bforce-comp}. The local maximum point is then estimated by interpolating the 
first derivative of these data. These maxima point are $Q^{CI}_c=0.81$, $Q^{MBA}_c=0.78$ and $Q^{HBA}_c=0.73$ 
(see Figure~\ref{fig:bforce-comp}), meaning that in the range $0.65\lessapprox Q\leq Q_c$ the heuristic 
strategies detach from the optimal point and may not show better results than an average over several 
random strategies, which shows an inflection point at $\sim 0.73$ (see Figure~\ref{fig:linregress}). 
On the other hand, for high values of modularity, above $Q_c$, the heuristic strategies starts once again 
to get closer to the optimal dismantling set and a thence more efficient than the expected value for 
random realizations.

It is important to note that due to the computational complexity of the optimal dismantling set problem, 
we had to restrict this study to $N=100$ nodes. This results in important finite size effects such as 
structural cutoffs of the degree distribution, impacting the overall network robustness. In this sense, 
the structural cutoff limits the maximum degree, which in turn restricts both the degree correlations 
and the range of modularity LFR models are able to generate \cite{Boguna2004}.
Another relevant discussion is the sensibility of our results to degree distribution. In a network 
with an arbitrary degree distribution, the critical fraction threshold of failures or targeted attacks 
for the existence of a giant component usually depends on the degree distribution~\cite{netscibook}. 
However, due to the network size limitations of this experimental study, one cannot measure precisely 
the sensibility of our experiment to the average degree distribution, and we must keep a fixed degree 
distribution in order to generate networks with varying modularity in an acceptable range. 
Therefore, we have kept a fixed average degree $<k>\sim3$, studying only the sensibility of the 
LFR benchmark networks to the modularity $Q$.

\section{Discussion}
Modular networks are known to be fragile in general, but since what values of modularity this weakness stands
out? To fill this gap we propose a computational experiment to check how heuristic methods are really close
to the optimum dismantling set. We do so by working with finite networks of size $N=100$ and checking all  
${100}\choose{5}$ $\approx 10^8$ possible removal lists of 5 nodes, due to computational limitations. 
Networks are generated according to the LFR algorithm, and have $k\sim 3$ and modularity $Q$ ranging from 
$0.59$ to $0.85$. The optimal point curve sets the lower limit to what is expected of the heuristic methods, 
we can see that as the modularity rises, the three middle curves approach the optimal values, the expected 
values curve also tends towards it, but much more slowly.

Firstly, results show the upper (expected) and lower (optimal) curves which represent the worst and the best 
scenarios in terms of damage to a network.

The behavior of those two curves with the modularity is very significant, showing a clear gap between 
the two, which spreads out when modularity increases.
As a general trend, it is evident that the damage of any kind of attack, even random ones, has a 
negative correlation with modularity, \emph{i.e} modularity is inversely related with robustness.
Indeed, not only MBA shows a better performance with modularity as expected, but HBA, and CI too. 
The reason is that the optimum attack (the black line and dot curve) shows a monotonic inverse 
dependence with $Q$.
However, for modularity $Q\lessapprox 0.7$ heuristic and random attacks are very similar, and the only way to 
inflict a more serious attack to a network with modularity in that range is by brute force, so we can say that 
a network in that range is relatively safe against random or malicious limited attacks ---considering the 
removal of only $5\%$ of nodes.

CI and MBA attacks are almost equivalent in efficiency (with MBA showing better performance), except for 
large values of modularity after a critical point ($Q^{CI}_c=0.81$ and $Q^{MBA}_c=0.78$). HBA is always 
the best heuristic method of attack, but for values of $Q > 0.7$ as we explained before, and goes very 
close to the optimum attack for large values of modularity ($Q^{HBA}_c=0.73$) but with high computational cost.

One should note, nonetheless, the many limitation of our study. As aforementioned, due to computational 
limitations, we could only remove $5$ out of $100$ nodes which even with the optimum set cannot fully stop 
the networks from having giant components. This same restraint implied that in order to study the sensibility 
of network robustness to modularity we had to keep the average degree constant and could not check the 
variability of our study to connectivity. Again, this limitation also implies finite size size effects such 
as high degree cutoff of the degree distribution, impacting network robustness. 

%% Havlin SCi Report
Recently, Wandelt \emph{et al.} presented a detailed and exhaustive
performance comparison between 13 heuristic methods of network attack
based on several criteria, including all the methods tested here.~\cite{wandelt2018comparative}
%({\B sera que eles testaram o nosso seguindo o procedimento completo?})
They came to a similar conclusion as the one we arrived in that HBA is
the best method in producing the greatest damage, not taking into
consideration the scalability of running time.
%Although, when considering computational cost, HBA is the only method
%that the running time scales with $10^3$ when the number of nodes is
%increased in one order of magnitude, making it computationaly
%unfeasible for large networks.
One of the best alternative methods of fragmentation ends up being the
approximated version of HBA, for the achieved results are similar with
much lower computational cost.

The authors also point out that wile HBA and its approximated version
obtain equivalent results for certain types of networks, the
correlation between the sequence of nodes that each method selects for
removal is low, pointing towards the existence of different ``paths"
to the optimal fragmentation.  This is one aspect we were able to
observe, because most networks used in our experiment do show many
different sequences of nodes that, when removed, dismantle the network
to the smallest giant component.

Due to the infeasibility that was early mentioned, Wandelt \emph{et
  al.}  could not compare any method with the optimal removal of
nodes.  In this sense, the brute force analysis of the relation
between heuristic methods and the optimal fragmentation point in a
varying modularity scenario is the main contribution of our study.

Moreover, we estimated an empirical critical point from which
heuristic methods get close enough to the optimal dismantling set,
serving as a guide to choosing the most efficient fragment strategy
for real modular networks. All in all, even considering the
limitations of this study, our experimental results support previous
theoretical researches that indicate intrinsic vulnerabilities of
modular networks, pointing out to modularity as the Achilles' heel of
real networks.

\bibliography{bruno_bib}

\end{document}